\documentclass[aps,prd,floatfix,nofootinbib,noshowpacs,preprint,10pt]{revtex4}

\usepackage[dvips]{graphicx}
\usepackage{epsfig}
\usepackage{float,epsfig}
\usepackage{dcolumn}
\usepackage{bm}
\usepackage{amsmath}
\usepackage{amssymb}

\def\p{\partial}                                
\def\slashchar#1{\setbox0=\hbox{$#1$}           
   \dimen0=\wd0                                     
   \setbox1=\hbox{/} \dimen1=\wd1                   
   \ifdim\dimen0>\dimen1                            
      \rlap{\hbox to \dimen0{\hfil/\hfil}}          
      #1                                            
   \else                                            
      \rlap{\hbox to \dimen1{\hfil$#1$\hfil}}       
      /                                             
   \fi}

\def\hc{{+{\rm h.c.}} }

\newcommand{\beq}{\begin{equation}}
\newcommand{\eeq}{\end{equation}}
\newcommand\be{\begin{equation} }
\newcommand\bea{\begin{eqnarray}}
\newcommand\ee{\end{equation}}
\newcommand\eea{\end{eqnarray}}

\DeclareMathOperator*{\str}{STr}

\def\endtitle{\par\end{quotation}\vskip3.5in minus2.3in\newpage}

\def\m{\mu}          
       \def\r{\rho}

\begin{document}
\title {The role of the Seiberg-Witten field redefinition in renormalization
of noncommutative chiral electrodynamics}

\author{Maja Buri\'c}%
\affiliation{%
University of Belgrade, Faculty of Physics \\ P.O.Box 44, RS-11001
Belgrade, Serbia
}%
\author{Du\v sko Latas}%
\affiliation{%
University of Belgrade, Faculty of Physics \\ P.O.Box 44, RS-11001
Belgrade, Serbia
}%
\author{Biljana Nikoli\' c}%
\affiliation{%
University of Belgrade, Faculty of Physics \\ P.O.Box 44, RS-11001
Belgrade, Serbia }%
\author{Voja Radovanovi\'c}%
\affiliation{%
University of Belgrade, Faculty of Physics \\ P.O.Box 44, RS-11001
Belgrade, Serbia
}%

\begin{abstract}
It has been conjectured in the literature that renormalizability
of the $\theta$-expanded noncommutative gauge theories improves
when one takes into account  full nonuniqueness of the Seiberg-Witten expansion,
which relates noncommutative (`high-energy') with commutative
(`low-energy') fields. In order to check this conjecture we analyze
renormalizability of the noncommutative chiral electrodynamics:
we quantize the action which contains all possible terms implied by the SW map.
After renormalization we arrive at a different theory in which the relation between the
coupling constants is changed. This means that the
$\theta$-expanded chiral electrodynamics is not  renormalizable:
 when  fermions are included, the
SW expansion is not preserved in quantization.
\end{abstract}


\maketitle

\section{Introduction}

Basic noncommutativity of spacetime coordinates~\cite{Snyder:1946qz}
is a very plausible idea when one ponders  two  singularity problems
of classical and quantum field theory: singular
 solutions and renormalizability. This can be easily seen:  when
coordinates are represented by operators $\hat x^\mu$,
commutation relations
\begin{equation}
[\hat x^\mu, \hat x^\nu] =\mathrm{i}\hat \theta^{\mu\nu}(\hat x)  ,         \qquad    \qquad \mu,\nu = 1,\dots,d
          \label{Nc}
\end{equation}
put  a lower bound of order $\,\vert\theta\vert^\frac{1}{2}$ on coordinate
measurements and an upper bound of order $\,\hbar\vert\theta\vert^{-\frac{1}{2}}$
on momentum measurements.
Here $\,\vert\theta\vert^\frac{1}{2}$
is the scale of noncommutativity  $\, \hat \theta^{\mu\nu}$.  This property
is desirable as a cure to both of the mentioned problems,
 but  it opens a whole range of questions, conceptual and concrete. First, one
 has  to define operationally a  `noncommutative
space' ${\cal A}\, $ which fulfills (\ref{Nc}), preferrably with a notion of smoothness
or differentiability, and second,  fields on it.  This for itself
is a difficult mathematical  problem. But a necessary constraint which one
wishes to impose is that noncommutative field theories have  good
commutative limit, the one which is experimentally  well established at the present
length scale, providing at the same time  resolution to the singularity
problems at the noncommutativity scale.

This of course is not an easy task. The most feasible model
of noncommutativity which we usually start with is the space
with constant, `canonical' noncommutativity:
\begin{equation}
[\hat x^\mu, \hat x^\nu] =\mathrm{i}\theta^{\mu\nu}={\rm const},      \label{Ncflat}
\end{equation}
because  fields  $\hat \phi$,  $\hat \chi $
on it can be represented by functions on ordinary $\mathbb{R}^4$.
The field multiplication is  given by the Moyal-Weyl star-product:
\begin{equation}
\hat\phi(x)\star \hat\chi(x) = \mathrm{e}^{\frac{\mathrm{i}}{2}
\theta^{\mu\nu}\frac{\partial}{\partial
x^\mu}\frac{\partial}{\partial y^\nu}}\hat\phi(x)\hat\chi(y)|_{y\to x} .
\label{moyal}
\end{equation}
This representation is called the Moyal space. The Moyal space is  a flat
noncomutative space, but clearly  in any number
of dimensions except in $d=2$,  constant commutator
(\ref{Ncflat})  breaks  the Lorentz symmetry.

Gauge symmetries on the Moyal space can  be, in principle, introduced
in a straightforward way. For example, for 
spinor field $\hat \psi$ one can define the action of
the noncommutative $\mathrm{U}(1)$ gauge group by
\begin{equation}
 \hat\psi^\prime = \hat U \hat\psi             ,           \label{hatpsi}
\end{equation}
the  $\hat U $ are unitary elements of  ${\cal A}$.
As coordinates $\hat x^\mu$ generate ${\cal A}$,
$\hat U$ are always expressible as functions,
$\hat U =\hat U (\hat x^\mu)$: we are dealing with local symmetry.
The group action can also be the adjoint,
$\,  \hat\psi^\prime = \hat U \hat\psi  \hat U^{-1}$
or the right action,
$\, \hat\psi^\prime= \hat\psi  \hat U^{-1}  $.
Obviously, the noncommutative $\mathrm{U}(1)$ group is not abelian,
and therefore transformation properties
of the gauge potential  resemble those of nonabelian  theories.
In fact, the expansion of  the potential in terms of the Lie algebra
generators, $ \, \hat A_\mu =\hat A_\mu^a T^a \, $,
is possible only for the $\mathrm{U}(N)$ groups. 
When one considers general noncommutative spaces,
quite often one can define only infinitesimal symmetry
transformations. 

In other aspects, also, noncommutative gauge symmetries exibit
new features. They are particularly
interesting when one considers  noncommutative spaces
defined as limits of  $N\times N$ matrix spaces
for $N\to\infty$. Then for example, the elements of the
noncommutative $\mathrm{U}(1)$ gauge
group, the unitary $N\times N$ matrices, belong at the same time
to the ordinary $\mathrm{U}(N)$: in a way,  the local noncommutative
$\mathrm{U}(1)$  can be identified with the $\mathrm{U}(\infty)$
on the commutative space. This connection can be extended to 
the Chern-Simons action, \cite{Polychronakos:2000nt}.
The  natural mixing of gauge and spatial degrees of freedom
in noncommutative geometry \cite{Szabo:2001kg},
can be further elaborated in the framework of matrix
models as a possibility to interpret gravity as emergent, that is, as the
 $\mathrm{U}(1)$ part of the $\mathrm{U}(N)$ symmetry group,
\cite{Steinacker:2010rh}.  These novel aspects  and
relations should be investigated and understood in more details.

In this paper we will be mainly concerned with gauge theories which include
fermions. We said that one of the important constraints on  
noncommutative theories is their commutative limit $\theta\to 0$.
In this limit one naturally expects that  noncommutative fields
$\hat A_\mu$, $\hat \psi$   reduce to  commutative gauge and matter fields
$A_\mu$ and $\psi$,
\begin{equation}
\hat A_\mu\vert_{\theta=0}= A_\mu, \qquad\hat\psi\vert_{\theta=0}=\psi,
\end{equation}
as  for $\, \theta=0$ the star product reduces to the ordinary one. We see
therefore that all noncommutative theories have the same commutative limit. 
On the other hand, starting from a specific commutative theory one can get 
various deformations: noncommutative generalization is not unique. But the
noncommutative structure of the space itself can give some restrictions
which reduce the number of possible models.

We mentioned that for symmetry transformations
defined by (\ref{hatpsi}) on the Moyal space,
only the $\mathrm{U}(N)$  groups
can be consistently represented. Various aspects of
this kind of models were analyzed in the literature~\cite{Minwalla:1999px,Hayakawa:1999yt,Matusis:2000jf,VanRaamsdonk:2000rr,Armoni:2000xr},
and it was shown that in perturbative quantization they
behave  worse than commutative gauge theories:
the ultraviolet  divergences  `propagate' to the infrared sector (UV/IR mixing).
Another widely explored possibility of representing
gauge symmetries  is the  enveloping algebra formalism, in which
one  enlarges the Lie algebra of the group to the enveloping algebra,
expanding  the gauge potential  in the symmetrized products of
the group generators $T^a$,
\begin{equation}
 \hat A_\mu =  \hat A_\mu^a T^a +  \hat A_\mu^{ab} \{T^a, T^b\}+\dots
\end{equation}
In this approach there is no restriction on the type of the gauge group,
but the UV/IR mixing  remains in 
straightforward quantization, \cite{Schupp:2008fs,Horvat:2011bs}.
In the original version of the enveloping-algebra formalism
\cite{Madore:2000en,Jurco:2001rq,Schupp:2002up},  one expands the theory in
$\theta$: the expansion of fields is called the Seiberg-Witten
(SW) map \cite{Seiberg:1999vs}. This expansion, apart from giving
the effective low-energy theory and the new interactions,
can in principle  be used to define the quantization procedure.
The idea is the following: one  first quantizes the theory in the first order,  then
proceeds to the second and   higher  orders by using some kind of iterative
procedure. Higher orders of fields in the SW expansion are
related to lower orders  by gauge symmetries  \cite{Ulker:2007fm,Aschieri:2011ng},
so one can hope that renormalizability will be achieved through
a noncommutative version of the Ward identities.  In addition,  the SW
expansion has an amount of nonuniqueness which increases the number of
possible counterterms for renormalization.

Therefore, in order to discuss the $\theta$-expanded theories we must first
 investigate their  behavior in  linear order. The results,
especially for the pure gauge theories, are quite encouraging.
The idea that the SW nonuniqueness \cite{Asakawa:1999cu}
can be used to obtain renormalizability was  proposed and used
first in~\cite{Bichl:2001cq} to show that the photon self-energy diagram
in the noncommutative $\mathrm{U}(1)$  theory is renormalizable to all orderds
in $\theta$. Linear-order action for the $\mathrm{SU}(N)$ gauge theory
was analyzed in~\cite{Buric:2005xe,Latas:2007eu} and found to be
renormalizable; a similar result was obtained for the gauge sector of a suitably defined
noncommutative Standard Model,~\cite{Buric:2006wm}.
Inclusion of the matter  on the other hand presents a difficulty.
For some time it was believed  that fermions cannot be successfully
incorporated into  a renormalizable theory because of the
4$\psi$ divergence,~\cite{Wulkenhaar:2001sq,Buric:2004he}; however,
this  divergence is absent in chiral theories,~\cite{Buric:2007ix}.
In fact,  we showed that all perturbative divergences
of noncommutative chiral electrodynamics are potentially removable
 by the Seiberg-Witten redefinition of  fields,~\cite{Buric:2010wd}.
Similar positive results
were obtained in \cite{Martin:2009sg,Martin:2009vg,Tamarit:2009iy}
for the GUT inspired anomaly-safe models with chiral fermions: it was found
 that all  linear-order divergences are given by marginal operators.
A counterargument  was given by Armoni~\cite{Armoni:2011pa} who, comparing
the expanded to the  unexpanded  theories, argues that the sum
of marginal operators of different orders gives in fact a relevant operator
which prevents renormalizability. To clarify the question of renormalizability and
decide whether the potential renormalizability of the $\theta$-expanded
 chiral electrodynamics obtained in~\cite{Buric:2010wd} is in fact
actual,  we undertake in this paper a task to renormalize
the model explicitely.  In order to do that we have to take
into account  the full nonuniqueness of the SW expansion, which
gives six new  interaction terms in the action.
The new coupling constants $\kappa_i$  however
are  constrained by one relation, (\ref{relation}).
We calculate all one-loop divergences and perform renormalization
of $\kappa_i$, including  noncommutativity  $\theta^{\mu\nu}$:
we obtain that  renormalization of the coupling constants
violates the constraint equation. Our  conclusion is: the SW
expansion cannot hold simultaneously for the bare and
the dressed fields.  This means that the model defined as 
 low-energy part of a basic noncommutative gauge
theory given by  (\ref{act})  and (\ref{sw}), is not renormalizable.

\section{Noncommutative chiral electrodynamics}

Let us introduce  the lagrangian. We wish to discuss  minimal
noncommutative extension of the commutative  chiral electrodynamics
defined by the action
\begin{equation}
S_{\rm C} = \int \mathrm{d}^4x \left( \mathrm{i}\bar\varphi \bar\sigma^\mu (\partial_\mu +\mathrm{i}qA_\mu )\varphi - \frac{1}{4}F_{\mu\nu}F^{\mu\nu} \right),
\label{Scom}
\end{equation}
where  $\varphi$ is a left chiral fermion of charge $q$, $A_\mu$ is
the vector potential and  $F_{\mu\nu} =\p_\mu A_\nu - \p_\nu A_\mu$ is the
electromagnetic field strength. Noncommutative fields
$\,\hat \varphi\,$, $\,\hat A_\mu\, $ and  $\hat F_{\mu\nu} = \partial_\mu \hat A_\nu - \partial_\nu \hat A_\mu +\mathrm{i}q[\hat A_\mu \stackrel{\star}{,} \hat A_\nu]$
 are also represented by functions of commutative coordinates;
 the corresponding   noncommutative action is given by
\begin{equation}
S_{\rm NC} = \int \mathrm{d}^4x \left( \mathrm{i}\hat{\bar \varphi} \star \bar\sigma^\mu (\partial_\mu +\mathrm{i}q\hat A_\mu ) \star \hat\varphi- \frac{1}{4}\hat F_{\mu\nu}\star \hat F^{\mu\nu}\right).
\label{act}
\end{equation}
Two sets of fields are related by the  Seiberg-Witten map:
\begin{equation}
\hat A_\rho =  A_\rho +\sum_{n=1}  A_\rho^{(n)}
 ,\qquad \hat \varphi = \varphi  +\sum_{n=1} \varphi^{(n)} .
\label{sw}
\end{equation}
This map is  an expansion  in powers of noncommutativity $\,\theta^{\mu\nu}$ which
 is designated by  index $(n)$: in the commutative limit 
higher powers vanish and  the initial values of  fields are $\,\varphi^{(0)}= \varphi $,   $\, A_\r^{(0)}= A_\r$.
Seiberg-Witten expansion (\ref{sw}) can be seen as  solution
to the condition that infinitesimal symmetry transformations close.
The simplest solution in linear order is given by~\cite{Madore:2000en,Jurco:2001rq,Schupp:2002up}:
\begin{eqnarray}
\hat A_\rho &=& A_\rho  +\frac{1}{4} q\, \theta^{\mu\nu} \{ A_\mu, \partial_\nu A_\rho   + F_{\nu \rho} \} , \label{expansion:A}
\\
\hat F_{\rho\sigma} &=& F_{\rho\sigma}  -\frac{1}{2}q\, \theta^{\mu\nu} \{ F_{\mu\rho} ,F_{\nu\sigma} \} \nonumber \\
& & + \frac{1}{4} q\, \theta^{\mu\nu} \{ A _\mu, (\partial_\nu + {D}_\nu )F_{\rho\sigma}  \} ,
\label{expansion:F}\\
\hat\varphi &=& \varphi +\frac{1}{2}q\, \theta^{\mu\nu} A_\mu \partial_\nu\varphi ,
\label{expansion:psi}
\end{eqnarray}
where ${D}_\mu$ denotes the commutative covariant derivative, $\, D_\m \varphi= (\partial_\mu +\mathrm{i}qA_\mu)\varphi  \, $.
It is  possible to generate  higher orders in the expansion
from  linear order, \cite{Ulker:2007fm}.   In addition, whenever we have a particular solution
$ A_\r^{(n)}$,  $ \varphi ^{(n)}$, we can obtain a more general one by
adding arbitrary gauge covariant expressions $ {\bf A}_\r^{(n)}$, ${\bf \Phi}^{(n)} $
of the same order~\cite{Asakawa:1999cu,Bichl:2001cq}:
\be
{A_\r^{\prime (n)}} =  A_\r^{(n)} + {\bf A}_\r^{(n)},\qquad
{\varphi^{\prime (n)}} = \varphi ^{(n)} + {\bf \Phi}^{(n)} .    \label{redef}
\ee
This property is called the Seiberg-Witten nonuniqueness:
it means that the relation
between the `physical',  high-energy fields $\, \hat A_\r$, $\hat \varphi\,$,
and the usual, experimentally observed low-energy fields
$\, A_\r$, $\varphi\,$ is not uniquely defined beyond the zeroth
order in $\theta$. One  intuitively expects that such a difference
would be unobservable. But in fact the  SW field redefinition
can change not only the dispersion relations or the cross
sections: it changes even  the renormalization properties of the theory, and
this happens when fermionic matter is included. Therefore, renormalizability
of the theory was proposed in the literature as a criterion which fixes the nonuniqueness
of the Seiberg-Witten expansion.

We shall discuss  $\theta$-linear order of the chiral electrodynamics.
Using  SW expansions (\ref{expansion:A}-\ref{expansion:psi}) we
obtain the action
\begin{equation}
\mathcal{L}_{\rm NC} = \mathcal{L}_{0,A}+\mathcal{L}_{0,\varphi} +\mathcal{L}_{1,A}+\mathcal{L}_{2}, \label{lag}
\end{equation}
with
\begin{eqnarray}
\mathcal{L}_{0,A}&=& - \frac{1}{4}F_{\mu\nu}F^{\mu\nu} \label{L0A}, \\[4pt]
\mathcal{L}_{0,\varphi} &=& \mathrm{i}\bar\varphi\bar\sigma^\mu( D_\mu \varphi ) \label{L0phi}, \\[4pt]
\mathcal{L}_{1,A}&=& \frac{1}{2}q\,\theta^{\mu\nu} \Big( F_{\mu\rho}F_{\nu\sigma}F^{\rho\sigma} -\frac{1}{4} F_{\mu\nu}F_{\rho\sigma}F^{\rho\sigma}\Big) , \label{L1A} \\
\mathcal{L}_{2} &=& - \frac{\rm i}{16}q\, \theta^{\mu\nu} \Delta^{\alpha\beta\gamma}_{\mu\nu\rho}
\, F_{\alpha\beta} \,\bar\varphi\,
\bar\sigma^\rho ( D_\gamma \varphi   ) \, \hc    ;     \label{L1phi}
\end{eqnarray}
we denote  in the following $\ \varepsilon^{\alpha\beta\gamma\delta}\varepsilon_{\mu\nu\rho\delta}
=- \Delta^{\alpha\beta\gamma}_{\mu\nu\rho}$.
Action (\ref{lag}) was analyzed in  \cite{Buric:2010wd} and it was shown that,
as it stands, it is not renormalizable. However,
 all divergences are of the form implied by a SW redefinition
and therefore we conjectured  that   there exists another
 expansion within allowed class   (\ref{redef})
which gives a renormalizable theory.

In order to check this conjecture, we need to expand the lagrangian
using  the most general first-order SW solution. Therefore
instead of (\ref{expansion:A}-\ref{expansion:psi}) we use
\be
{A_\r^{\prime }} =  A_\r+ {\bf A}_\r^{(1)},\qquad
{\varphi^{\prime }} = \varphi + {\bf \Phi}^{(1)} ,     \label{redef1}
\ee
where ${\bf A}_\r^{(1)} $ and ${\bf \Phi}^{(1)} $ are
covariant expressions of  first order in $\theta $:
\begin{eqnarray}
{\bf A}^{(1)}_\rho &= & a_1 \theta^{\mu\nu} \varepsilon_{\mu\rho\sigma\tau}(\partial_\nu F^{\sigma\tau}) + a_2 \theta^{\mu\nu}\varepsilon_{\mu\nu\rho\tau}(\partial_\sigma F^{\tau\sigma}) \nonumber \\
  & & + a_3  \theta^{\mu\nu}\varepsilon_{\mu\nu\tau\sigma}(\partial_\rho F^{\tau\sigma}),   \label{a} \\[4pt]
{\bf \Phi}^{(1)} &=&\mathrm{i} b_1 \theta^{\mu\nu} \sigma_{\mu\nu}(D^2 \varphi) + b_2 {q}\theta^{\mu\nu}   {F_{\mu}}^{\rho} \sigma_{\nu\rho }\varphi \nonumber \\
 & & + b_3 {q}\theta^{\mu\nu}F_{\mu\nu}\varphi + \mathrm{i} b_4 {q}\theta^{\mu\nu} \varepsilon_{\mu\nu\rho\sigma} F^{\rho\sigma}\varphi, \label{b}
\end{eqnarray}
and constants $a_{i}$ and $b_{i}$ are real.
This changes the initial action to
\begin{equation}
 S_{\rm NC}^\prime = S_{\rm NC} +\Delta S^{(1)}_{\rm SW},
\end{equation}
with
\begin{eqnarray}
\Delta S^{(1)}_{\rm SW} &=&\int {\rm d}^4 x \left( (D_\rho F^{\rho\mu}){\bf A}_\mu^{(1)}- q \bar\varphi\bar\sigma^\mu\varphi \mathbf{A}_\mu^{(1)}\right. \nonumber \\
& & + \left.\left(\mathrm{i}\bar\varphi \bar\sigma^\mu(D_\mu {\bf \Phi}^{(1)})\hc\right) \right).
\end{eqnarray}
When we introduce (\ref{a}-\ref{b}) and simplify the action using various identities,
we obtain
\begin{widetext}
\begin{eqnarray*}
\Delta \mathcal{L}^{(1)}_{\rm SW}&=&  \mathrm{i}\frac{b_{1}}{2}  \theta^{\mu\nu}\varepsilon_{\mu\nu\rho\sigma} \bar\varphi \bar\sigma^{\sigma} D^{\rho} D^{2} \varphi \\
& & + q\theta^{\mu\nu}\left[ -\mathrm{i}\left(b_{1}+\frac{b_{2}}{2}\right) F_{\mu\rho} \bar\varphi \bar \sigma_{\nu} D^{\rho}\varphi + \mathrm{i}\frac{b_{2}}{2} F_{\mu\rho} \bar\varphi \bar \sigma^{\rho} D_{\nu}\varphi + \mathrm{i}b_{3} F_{\mu\nu} \bar\varphi \bar \sigma^{\rho} D_{\rho}\varphi\right.\\
& & + \left.\left(a_{1}+a_{2} -\frac{b_{2}}{4}\right) \varepsilon_{\mu\rho\sigma\tau}F^{\rho\sigma} \bar\varphi \bar \sigma^{\tau} D_{\nu}\varphi + \left(a_{3} -\frac{a_{2}}{2} +b_{4}\right) \varepsilon_{\mu\nu\rho\sigma}F^{\rho\sigma} \bar\varphi \bar \sigma^{\tau} D_{\tau}\varphi\right] \hc  .
\end{eqnarray*}
\end{widetext}
This form is in a way canonical  as it contains minimal number of terms.
The new lagrangian, our starting point for quantization, reads
\begin{equation}
\label{PolazniLag}
{\mathcal{L}}^\prime_{\rm NC} = \mathcal{L}_{\rm C} + \mathcal{L}_{1,A} + (1+\kappa_2) \mathcal{L}_{2} +  \sum_{i=3}^7 \kappa_i \mathcal{L}_{i},
\end{equation}
where $\kappa_i$, $i=2,\dots,7$ are the coupling constants introduced as
\begin{equation}
 \kappa_2=-b_2,\quad \kappa_3=\frac{b_1}{2},\quad \kappa_4=-b_1-\frac{b_2}{2},
\label{**}
\end{equation}
\begin{equation}
 \kappa_5=\frac{b_2}{4}+b_3,\quad \kappa_6=a_1+a_2-\frac{b_2}{4},\quad \kappa_7=a_3-\frac{a_2}{2}+b_4,
\label{***}
\end{equation}
and
\begin{eqnarray}
\mathcal{L}_{3}&=&{\mathrm{i}} \theta^{\mu\nu}\varepsilon_{\mu\nu\rho\sigma} \bar\varphi \bar\sigma^{\sigma} D^{\rho} D^{2} \varphi\hc \label{L3} \\[4pt]
\mathcal{L}_{4}&=& \mathrm{i} q\theta^{\mu\nu} F_{\mu\rho} \bar\varphi \bar \sigma_{\nu} D^{\rho}\varphi \hc \label{L4} \\[4pt]
\mathcal{L}_{5}&=& \mathrm{i} q\theta^{\mu\nu} F_{\mu\nu} \bar\varphi \bar \sigma^{\rho} D_{\rho}\varphi \hc \label{L5} \\[4pt]
\mathcal{L}_{6}&=& q\theta^{\mu\nu} \varepsilon_{\mu\rho\sigma\tau}F^{\rho\sigma} \bar\varphi \bar \sigma^{\tau} D_{\nu}\varphi \hc \label{L6} \\[4pt]
\mathcal{L}_{7}&=& q\theta^{\mu\nu} \varepsilon_{\mu\nu\rho\sigma}F^{\rho\sigma} \bar\varphi \bar \sigma^{\tau} D_{\tau}\varphi \hc \label{L7}  .
\end{eqnarray}
Note that not all coupling constants are independent: there is a relation between them,
\begin{equation}
 \kappa_2-4\kappa_3-2\kappa_4 =0              .               \label{relation}
\end{equation}
We shall see that this relation is broken in quantization.

\section{Quantization}

For quantization we use the background field method. The procedure for
this kind of a model was developed in details
in \cite{Buric:2010wd} so we will not repeat it
here. The main difference is that now, instead of one, we have six fermion-photon vertices;
 in addition, the fermion propagator has noncommutative correction which we
also treat perturbatively. For the purpose of calculation
of  functional integrals we   introduce the Majorana spinor $\psi$
instead of the chiral $\varphi$,
$\
\psi = \left(
    \begin{array}{c}
        \varphi_\alpha \\
       \bar\varphi^{{\dot\alpha}}
    \end{array}
\right) .
$
Rewriting the  action in Majorana
spinors, for the commutative part of the spinor lagrangian we obtain
$\  \mathcal{L}_{0,\varphi}  =
\frac{\mathrm{i}}{2}\,\bar\psi\gamma^\mu(\partial_\mu
-\mathrm{i}q\gamma_5 A_\mu )\psi  ;       \   $
noncommutative terms are expressed likewise.

The one-loop effective action is given by expansion
\begin{widetext}
\begin{eqnarray}
\Gamma^{(1)} & =& \frac{\mathrm{i}}{2}\, \mathrm{STr} \log \left(\mathcal{I} +\Box^{-1}N_1 +\Box^{-1}T_0 +\Box^{-1}T_1+\Box^{-1}T_2\right) \\
&=&\frac{\mathrm{i}}{2} \sum\frac{(-1)^{n+1}}{n} \mathrm{STr}\left(\Box^{-1}N_1 +\Box^{-1}T_0 +\Box^{-1}T_1 +\Box^{-1}T_2 \right)^n . \nonumber
\label{perturb}
\end{eqnarray}
\end{widetext}
In this formula, $N_1$ denotes a matrix which is obtained from  commutative 3-vertices
 after the  expansion of  quantum fields around the stationary
classical configuration. It is given by
\begin{equation}
N_1 =
q\begin{pmatrix}
   0 & \mathrm{i}\bar\psi\gamma_5\gamma^\kappa \slashchar{\partial} \\[4pt]
       -\gamma_5\gamma^\lambda\psi & \mathrm{i}\gamma_5\slashchar{A} \slashchar{\partial}
\end{pmatrix}.\nonumber
\end{equation}
The $T_0$, $T_1$ and $T_2$ are
 defined in analogy but related to the terms
 linear in $\theta$: $T_0$ corresponds to the 2-vertex, that is,
to the noncommutative correction of the fermion propagator,
while $T_1$ and $T_2$ are obtained from
3- and 4-vertices. The expansion  gives
\begin{equation}
T_0 =
2 \kappa_3 \theta^{\mu\nu}\varepsilon_{\mu\nu\rho\sigma}
    \begin{pmatrix}
       0 & 0 \\[4pt]
       0 & \gamma^\sigma\slashchar{\partial} \partial^\rho \Box
    \end{pmatrix}                                    .          \nonumber
\end{equation}
$T_1$  is the sum of six terms,
$\
T_1=\sum_{i=2}^7 T_1^{\kappa_i},\nonumber
$
with
\begin{widetext}
\begin{eqnarray}
T_1^{\kappa_2} &=&
\frac{1}{8}\kappa_2 q \theta^{\mu\nu}
 \Delta^{\alpha\beta\gamma}_{\mu\nu\rho}
      \begin{pmatrix}
      0 & 2\delta^\kappa_\alpha (\partial_\beta \bar\psi)\gamma^\rho\partial_\gamma\slashchar{\partial} \\[4pt]
      2{\mathrm{i}}\delta^\lambda_\alpha \gamma^\rho(\partial_\beta \psi)\partial_\gamma & F_{\alpha\beta}\gamma^\rho \partial_\gamma \slashchar{\partial} \\
 \end{pmatrix},\nonumber \\[8pt]
T_1^{\kappa_3} &=&
2 \kappa_3 q  \theta^{\mu\nu}\varepsilon_{\mu\nu\rho\sigma}
    \begin{pmatrix}
       0 & \mathrm{i} (g^{\rho\kappa} \bar\psi\Box - (\partial^\rho \bar\psi) \partial^\kappa + \bar\psi \partial^\rho \partial^\kappa ) \gamma^5 \gamma^\sigma \slashchar{\partial} \\[4pt]
       \gamma^5 \gamma^\sigma (g^{\rho\lambda} (\Box \psi) - \overleftarrow \partial^\rho (\partial^\lambda \psi) + (\partial^\rho {\partial^\lambda} \psi) ) & \mathrm{i} (A^\rho \Box  - \overleftarrow \partial^\rho A^\tau \partial_\tau + A_\tau \partial^\rho \partial^\tau)\gamma^5 \gamma^\sigma  \slashchar{\partial}
    \end{pmatrix}, \nonumber \\[8pt]
T_1^{\kappa_4} &=&
2 \kappa_4 q  \theta^{\mu\nu}
    \begin{pmatrix}
       0 &  (\overleftarrow{\partial}_\mu g_{\rho\kappa}-\overleftarrow{\partial}_\rho g_{\mu\kappa})\bar\psi \gamma_\nu \partial^\rho \slashchar{\partial} \\[4pt]
       \mathrm{i} \gamma_\nu (\partial^\rho \psi) (g_{\rho\lambda}\partial_\mu-g_{\mu\lambda}\partial_\rho) &  F_{\mu\rho}\gamma_\nu \partial^\rho \slashchar{\partial}
    \end{pmatrix}, \nonumber \\[8pt]
T_1^{\kappa_5} &=&
2  \kappa_5 q  \theta^{\mu\nu}
    \begin{pmatrix}
       0 & 2 \overleftarrow{\partial}_\mu g_{\nu\kappa}\bar\psi \Box \\[4pt]
       2\mathrm{i} \gamma^\rho (\partial_\rho \psi) g_{\nu\lambda}\partial_\mu &  F_{\mu\nu}\Box
    \end{pmatrix}, \nonumber \\[8pt]
T_1^{\kappa_6} &=&
2 \kappa_6 q  \theta^{\mu\nu} \varepsilon_{\mu\rho\sigma\tau}
    \begin{pmatrix}
       0 & 2\mathrm{i}  \overleftarrow{\partial}^\rho \delta_\kappa^\sigma \bar\psi \gamma^\tau \gamma_5 \partial_\nu \slashchar{\partial} \\[4pt]
       -2 \delta_\lambda^\sigma \gamma^\tau \gamma_5 (\partial_\nu\psi) \partial^\rho & \mathrm{i} F^{\rho\sigma}\gamma^\tau \gamma_5 \partial_\nu \slashchar{\partial}
    \end{pmatrix}, \nonumber \\[8pt]
T_1^{\kappa_7} &=&
2 \kappa_7 q  \theta^{\mu\nu} \varepsilon_{\mu\nu\rho\sigma}
    \begin{pmatrix}
       0 & -2\mathrm{i}  \overleftarrow{\partial}^\rho \delta^\sigma_\kappa\bar\psi \gamma^5 \Box \\[4pt]
       -2 \delta^\sigma_\lambda \gamma^\tau \gamma_5 (\partial_\tau \psi) \partial^\rho & -\mathrm{i} F^{\rho\sigma} \gamma_5\Box
    \end{pmatrix}  , \nonumber
\end{eqnarray}
and from  4-vertices we obtain that
$\
T_2=\sum_{i=2}^5 T_2^{\kappa_i}\,
$
with
\begin{eqnarray}
T_2^{\kappa_2} &=&
\frac{1}{8} \kappa_2 q^2  \theta^{\mu\nu} \Delta^{\alpha\beta\gamma}_{\mu\nu\rho}
    \begin{pmatrix}
        2\delta^\lambda_\beta \delta^\kappa_\gamma \bar\psi\gamma^\rho \gamma_5 \psi \partial_\alpha  & -\mathrm{i}\delta^\kappa_\gamma F_{\alpha\beta}\bar\psi \gamma^\rho \gamma^5 \slashchar{\partial} -2\mathrm{i}\overleftarrow{\partial}_\alpha \delta^\kappa_\beta  A_\gamma \bar\psi \gamma^\rho \gamma^5 \slashchar{\partial} \\[4pt]
        F_{\alpha\beta} \delta^\lambda_\gamma \gamma^\rho \gamma^5\psi + 2 A_\gamma \delta^\lambda_\beta \gamma^\rho \gamma^5 \psi  {\partial}_\alpha  &
        -\mathrm{i} F_{\alpha\beta}A_\gamma\gamma^\rho\gamma_5 \slashchar{\partial}
    \end{pmatrix}, \nonumber \\[8pt]
T_2^{\kappa_3} &=&
2 \kappa_3 q^2  \theta^{\mu\nu}\varepsilon_{\mu\nu\rho\sigma}
    \begin{pmatrix}
       - \mathrm{i} (2   \delta^\rho_\lambda \bar\psi\gamma^\sigma (\partial_\kappa \psi) +  g_{\lambda\kappa} \bar\psi\gamma^\sigma (\partial^\rho \psi)) &  -2 (A^\rho \bar\psi \partial_\kappa  + A^\tau \delta^\rho_\kappa \bar\psi \partial_\tau + A_\kappa \bar\psi \partial^\rho) \gamma^\sigma \slashchar{\partial} \\[4pt]
       -2 \mathrm{i} \gamma^\sigma(A^\rho(\partial_\lambda\psi) + \delta^\rho_\lambda A^\tau (\partial_\tau\psi) + A_\lambda (\partial^\rho\psi)) & - \gamma^\sigma (2 A^\tau A^\rho \partial_\tau +  A^\tau A_\tau \partial^\rho ) \slashchar{\partial}
    \end{pmatrix}, \nonumber \\[8pt]
T_2^{\kappa_4} &=&
2 \kappa_4 q^2  \theta^{\mu\nu}
    \begin{pmatrix}
        -g^{\rho\kappa} \bar\psi\gamma_5 \gamma_\nu \psi (\delta_\rho^\lambda \partial_\mu - \delta_\mu^\lambda \partial_\rho) & \mathrm{i}(F_{\mu\kappa} +(\overleftarrow{\partial}_\mu g_{\kappa\rho} - \overleftarrow{\partial}_\rho g_{\kappa\mu}) A^\rho) \bar\psi \gamma_5 \gamma_\nu \slashchar{\partial} \\[4pt]
        -\gamma_5\gamma_\nu\psi(F_{\mu\lambda}+ A^\rho ( g_{\lambda\rho} {\partial}_\mu -  g_{\lambda\mu}{\partial}_\rho)) &  \mathrm{i} F_{\mu\rho}A^\rho\gamma_5 \gamma_\nu\slashchar{\partial}
    \end{pmatrix}, \nonumber \\[8pt]
T_2^{\kappa_5} &=&
2 \kappa_5 q^2  \theta^{\mu\nu}
    \begin{pmatrix}
        2\delta_\mu^\lambda \bar\psi \gamma_5 \gamma^\kappa\psi \partial_\nu  & \mathrm{i}(F_{\mu\nu}\bar\psi \gamma_5 \gamma^\kappa +2\overleftarrow{\partial}_\mu \delta^\kappa_\nu  A^\rho \bar\psi \gamma_5\gamma_\rho) \slashchar{\partial} \\[4pt]
        -F_{\mu\nu}\gamma_5\gamma^\lambda\psi + 2 A^\rho \gamma^5 \gamma_\rho \psi  \delta^\lambda_\nu {\partial}_\mu  &
        \mathrm{i} F_{\mu\nu}\gamma_5 \slashchar{A} \slashchar{\partial}
    \end{pmatrix}. \nonumber
\end{eqnarray}
These operators are the basic ingredients for the perturbation theory.

\section{Divergences and renormalization}

By the power counting one can fix the terms which give divergent
contributions. They come from
\begin{eqnarray}
 {\Gamma}^{(1)}_{\rm div}&=&\frac{\rm i}{2} \str \left.\left( -\frac{1}{2}(\Box^{-1}N_1\Box^{-1}N_1) + \frac{1}{3} (\Box^{-1}N_1 \Box^{-1}N_1 \Box^{-1}N_1) \right. \right. \\
& & \left. \left.\phantom{\frac12} -(\Box^{-1}N_1\Box^{-1}T_1)-(\Box^{-1}N_1\Box^{-1}N_1\Box^{-1}N_1\Box^{-1}T_0)+(\Box^{-1}N_1\Box^{-1}N_1\Box^{-1}T_1)-(\Box^{-1}N_1\Box^{-1}T_2)\right)\right|_{\rm div}          . \nonumber
\end{eqnarray}
We have calculated  divergences
by dimensional regularization. In this
very demanding calculation our main aid, apart from
the {\it  Mathematica} based package {\it MathTensor},
was  the gauge covariance of the background field method.
Omitting  the intermediate steps,  we write  the final result:
\begin{eqnarray}
{\Gamma}^{(1)}_{\rm div}
&=& \frac{1}{(4\pi)^2 \epsilon} q^2 \int {\rm d}^4 x \left( {\rm i}\bar\psi \gamma^\mu (\partial_\mu - {\rm i} q \gamma_5 A_\mu) \psi -\frac{1}{3}F_{\mu\nu}F^{\mu\nu} \right. \label{Div}\\[4pt]
& & + \theta^{\mu\nu} \left. \left( \frac{\rm i}{12} {\varepsilon_{\mu\nu\rho\sigma}} \bar\psi \gamma^\sigma D^\rho D^\tau D_\tau \psi +q\left( \frac{2}{3} F_{\mu\rho}F_{\nu\sigma} F^{\rho\sigma} - \frac{1}{6} F_{\mu\nu}F_{\rho\sigma} F^{\rho\sigma} -\frac{5\,\rm i}{6} F_{\mu\rho} \bar\psi \gamma^\rho D_\nu\psi \right. \right.\right.\nonumber \\[4pt]
\nonumber
& & + \frac{\rm i}{6} F_{\mu\rho} \bar\psi \gamma_\nu D^\rho\psi + \frac{2\,\rm i}{3} F_{\mu\nu} \bar\psi \gamma^\rho D_\rho\psi\left.\left.+\frac{1}{6} {\varepsilon_{\mu\rho\sigma\tau}} F^{\rho\sigma} \bar\psi \gamma_5 \gamma^\tau D_\nu\psi -\frac{7}{8}  {\varepsilon_{\mu\nu\rho\sigma}} F^{\rho\sigma} \bar\psi \gamma_5 \gamma^\tau D_\tau\psi \right)\right. \nonumber \\[4pt]
\nonumber
& & +\kappa_2 \left( \frac{\rm i}{12} {\varepsilon_{\mu\nu\rho\sigma}} \bar\psi \gamma^\sigma D^\rho D^\tau D_\tau \psi +q\left( \frac{2}{3} F_{\mu\rho}F_{\nu\sigma} F^{\rho\sigma} - \frac{1}{6} F_{\mu\nu}F_{\rho\sigma} F^{\rho\sigma} +\frac{\rm i}{2} F_{\mu\rho} \bar\psi \gamma^\rho D_\nu\psi \right. \right.\\[4pt]
\nonumber
& & \left.\left.- \frac{3\,\rm i}{2} F_{\mu\rho} \bar\psi \gamma_\nu D^\rho\psi +\frac{5}{36} {\varepsilon_{\mu\rho\sigma\tau}} F^{\rho\sigma} \bar\psi \gamma_5 \gamma^\tau D_\nu\psi - \frac{1}{8} {\varepsilon_{\mu\nu\rho\sigma}} F^{\rho\sigma} \bar\psi \gamma_5 \gamma^\tau D_\tau\psi \right)\right) \\
\nonumber
& & +\kappa_3 \left( -\frac{4\, \rm i}{3} {\varepsilon_{\mu\nu\rho\sigma}} \bar\psi \gamma^\sigma D^\rho D^\tau D_\tau \psi +q\left(- \frac{16}{3} F_{\mu\rho}F_{\nu\sigma} F^{\rho\sigma} + \frac{4}{3} F_{\mu\nu}F_{\rho\sigma} F^{\rho\sigma}-\frac{10\,\rm i}{3} F_{\mu\rho} \bar\psi \gamma^\rho D_\nu\psi \right. \right.\\[4pt]
\nonumber
& & - \frac{34\,\rm i}{3} F_{\mu\rho} \bar\psi \gamma_\nu D^\rho\psi + \frac{20\,\rm i}{3} F_{\mu\nu} \bar\psi \gamma^\rho D_\rho\psi\left.\left.-\frac{11}{3} {\varepsilon_{\mu\rho\sigma\tau}} F^{\rho\sigma} \bar\psi \gamma_5 \gamma^\tau D_\nu\psi +2 {\varepsilon_{\mu\nu\rho\sigma}} F^{\rho\sigma} \bar\psi \gamma_5 \gamma^\tau D_\tau\psi \right)\right) \\[4pt] 
\nonumber
& & +\kappa_4 \left( -\frac{7\, \rm i}{6} {\varepsilon_{\mu\nu\rho\sigma}} \bar\psi \gamma^\sigma D^\rho D^\tau D_\tau \psi +q\left(- \frac{8}{3} F_{\mu\rho}F_{\nu\sigma} F^{\rho\sigma} +\frac{2}{3} F_{\mu\nu}F_{\rho\sigma} F^{\rho\sigma} -\frac{10\,\rm i}{3} F_{\mu\rho} \bar\psi \gamma^\rho D_\nu\psi \right. \right.\\[4pt]
\nonumber
& & - \frac{13\,\rm i}{3} F_{\mu\rho} \bar\psi \gamma_\nu D^\rho\psi + \frac{19\,\rm i}{6} F_{\mu\nu} \bar\psi \gamma^\rho D_\rho\psi\left.\left.+\frac{2}{3} {\varepsilon_{\mu\rho\sigma\tau}} F^{\rho\sigma} \bar\psi \gamma_5 \gamma^\tau D_\nu\psi
- \frac{1}{4} {\varepsilon_{\mu\nu\rho\sigma}} F^{\rho\sigma} \bar\psi \gamma_5 \gamma^\tau D_\tau\psi \right)\right) \\[4pt]
\nonumber
& & +\kappa_5 \left( \frac{\rm i}{3} {\varepsilon_{\mu\nu\rho\sigma}} \bar\psi \gamma^\sigma D^\rho D^\tau D_\tau \psi \right.
+q\left(-\frac{4\,\rm i}{3} F_{\mu\rho} \bar\psi \gamma^\rho D_\nu\psi
+ \frac{2\,\rm i}{3} F_{\mu\rho} \bar\psi \gamma_\nu D^\rho\psi + \frac{5\,\rm i}{3} F_{\mu\nu} \bar\psi \gamma^\rho D_\rho\psi \right.\\[4pt]
\nonumber
& & \left.\left.+\frac{2}{3} {\varepsilon_{\mu\rho\sigma\tau}} F^{\rho\sigma} \bar\psi \gamma_5 \gamma^\tau D_\nu\psi
- \frac{1}{2} {\varepsilon_{\mu\nu\rho\sigma}} F^{\rho\sigma} \bar\psi \gamma_5 \gamma^\tau D_\tau\psi \right)\right)\\[4pt]
\nonumber
& & +\kappa_6 \left(2 q {\varepsilon_{\mu\rho\sigma\tau}} F^{\rho\sigma} \bar\psi \gamma_5 \gamma^\tau D_\nu\psi\right) \\[4pt]
\nonumber
& & +\kappa_7 \left( -\frac{4\,\rm i}{3} {\varepsilon_{\mu\nu\rho\sigma}} \bar\psi \gamma^\sigma D^\rho D^\tau D_\tau \psi \right.
+q\left(\frac{4\,\rm i}{3} F_{\mu\rho} \bar\psi \gamma^\rho D_\nu\psi + \frac{4\,\rm i}{3} F_{\mu\rho} \bar\psi \gamma_\nu D^\rho\psi - \frac{8\,\rm i}{3} F_{\mu\nu} \bar\psi \gamma^\rho D_\rho\psi \right.\\[4pt]
\nonumber
& & \left.\left.\left.\left.-\frac{2}{3} {\varepsilon_{\mu\rho\sigma\tau}} F^{\rho\sigma} \bar\psi \gamma_5 \gamma^\tau D_\nu\psi
+ 2 {\varepsilon_{\mu\nu\rho\sigma}} F^{\rho\sigma} \bar\psi \gamma_5 \gamma^\tau D_\tau\psi \right)\right)\right)\right)    .
\end{eqnarray}
This result reduces to the one found before in \cite{Buric:2010wd}  for $\, \kappa_i=0$.
One  immediately  notices that,
apart from the usual commutative divergences and the
3-photon term $\mathcal{L}_{1,A}$, all other
terms are electron-photon interactions: they have
be transformed and expressed via $ \mathcal{L}_{i}$.

As usual, we write the one-loop divergent part  as
\begin{equation}
     {\Gamma}^{(1)}_{\rm div} = - {\mathcal{L}}^\prime_{\rm ct} ,
\end{equation}
and add counterterms ${\mathcal{L}}^\prime_{\rm ct} $ to the
classical action to cancel divergences after quantization.
In this manner we obtain the bare lagrangian:
\begin{eqnarray}
{\mathcal{L}}^\prime_{\rm NC}+{\mathcal{L}}^\prime_{\rm ct}&=& - \frac{1}{4}F_{\mu\nu}F^{\mu\nu}\left(1-\frac{4}{3} \frac{q^2}{(4\pi)^2 \epsilon}\right) + \frac{1}{2} \left(\mathrm{i}\bar\varphi\bar\sigma^\mu( D_\mu \varphi )\hc\right)  \left(1-2 \frac{q^2}{(4\pi)^2 \epsilon}\right) \label {c.t.} \\[4pt]
& & +\frac{1}{2}q \mu^{\frac{\epsilon}{2}} \theta^{\mu\nu} \left( F_{\mu\rho}F_{\nu\sigma}F^{\rho\sigma} -\frac{1}{4} F_{\mu\nu}F_{\rho\sigma}F^{\rho\sigma}\right) \left(1 + \frac{q^2}{(4\pi)^2\epsilon}\frac{-4(1 +\kappa_2 - 8\kappa_3 -4 \kappa_4 )}{3}  \right)\nonumber \\[4pt]
& & +\frac{1}{16}q \mu^{\frac{\epsilon}{2}} \theta^{\mu\nu} \Delta^{\alpha\beta\gamma}_{\mu\nu\rho}\, F_{\alpha\beta} \left({\rm i}\bar\varphi \bar\sigma^\rho ( D_\gamma \varphi   ) \hc\right)\left(1+ \kappa_2 +\frac{q^2}{(4\pi)^2 \epsilon} \frac{-5+3\kappa_2 - 20 \kappa_3 - 20 \kappa_4 - 8\kappa_5 + 8\kappa_7}{3}  \right) \nonumber \\[4pt]
& & + \left({\mathrm{i}} \theta^{\mu\nu}\varepsilon_{\mu\nu\rho\sigma} \bar\varphi \bar\sigma^{\sigma} D^{\rho} D^{2} \varphi\hc \right) \left( \kappa_3 + \frac{q^2}{(4\pi)^2 \epsilon} \frac{-1 -\kappa_2 + 16 \kappa_3 +14 \kappa_4 -4\kappa_5 +16\kappa_7}{12} \right)  \nonumber  \\[4pt]
& & +q\mu^{\frac{\epsilon}{2}}\left( \mathrm{i}\theta^{\mu\nu} F_{\mu\rho} \bar\varphi \bar \sigma_{\nu} D^{\rho}\varphi \hc\right) \left( \kappa_4 + \frac{q^2}{(4\pi)^2 \epsilon} \frac{-1 + 9\kappa_2+ 68 \kappa_3 + 26 \kappa_4 - 4\kappa_5 - 8 \kappa_7}{6} \right)\nonumber \\[4pt]
& & +q\mu^{\frac{\epsilon}{2}} \left( \mathrm{i}\theta^{\mu\nu} F_{\mu\nu} \bar\varphi \bar \sigma^{\rho} D_{\rho}\varphi \hc \right) \left( \kappa_5 + \frac{q^2}{(4\pi)^2 \epsilon} \frac{-1 - \kappa_2 - 20 \kappa_3 -6\kappa_4 - 4 \kappa_5 + 8 \kappa_7}{4} \right)\nonumber \\[4pt]
& & +q\mu^{\frac{\epsilon}{2}}\left(\theta^{\mu\nu} \varepsilon_{\mu\rho\sigma\tau}F^{\rho\sigma} \bar\varphi \bar \sigma^{\tau} D_{\nu}\varphi \hc \right) \left( \kappa_6 + \frac{q^2}{(4\pi)^2 \epsilon} \frac{-6 -5 \kappa_2+ 132 \kappa_3 -24 \kappa_4- 24 \kappa_5 - 72 \kappa_6 + 24 \kappa_7}{36} \right)\nonumber \\[4pt]
& & +q\mu^{\frac{\epsilon}{2}}\left(\theta^{\mu\nu} \varepsilon_{\mu\nu\rho\sigma}F^{\rho\sigma} \bar\varphi \bar \sigma^{\tau} D_{\tau}\varphi \hc\right) \left( \kappa_7 + \frac{q^2}{(4\pi)^2 \epsilon} \frac{ 7 +\kappa_2 -16\kappa_3 + 2\kappa_4 + 4\kappa_5 - 16 \kappa_7}{8}\right).\nonumber
\end{eqnarray}
\end{widetext}
The  $\epsilon$ is  regularization parameter  from
dimensional regularization. Now we can
 read off the values of the bare couplings and fields. From  the
commutative  part  we obtain  known renormalizations
\begin{eqnarray}
\varphi_{0}&=&\sqrt{Z_{2}}\, \varphi=\sqrt{1-2 \frac{q^2}{(4\pi)^2 \epsilon}}\, \varphi,  \label{11}\\[4pt]
A^\mu_0&=&\sqrt{Z_3}\, A^\mu=\sqrt{1-\frac{4}{3} \frac{q^2}{(4\pi)^2 \epsilon}}\, A^\mu, \\[4pt]
q_{0}&=&\mu^{\frac{\epsilon}{2}} Z_3^{-1/2}Z_{2}^{-1}   \left(1-2 \frac{q^2}{(4\pi)^2 \epsilon}\right) q \nonumber \\
&=&\mu^{\frac{\epsilon}{2}} \left(1+\frac{2}{3} \frac{q^2}{(4\pi)^2 \epsilon}\right) q.
\end{eqnarray}
The noncommutative part of divergences will give the bare couplings, $(\kappa_i)_0$.
But we see that along with $\mathcal{L}_{i} $, 3-photon term $\mathcal{L}_{1,A} $
also gets  quantum correction from the fermion loops, though
its coefficient is  in the classical lagrangian  fixed to 1. This implies that a rescaling of
  noncommutativity parameter $ \theta$ is  necessary. The bare $\theta_0$ is given by
\begin{equation}
 \theta^{\mu\nu}_0=\left(1-\frac{4}{3}\frac{q^2}{(4\pi)^2 \epsilon} (\kappa_2 - 8\kappa_3 - 4\kappa_4) \right) \theta^{\mu\nu} .                              \label{theta0}
\end{equation}
Using (\ref{11}-\ref{theta0})  we  obtain  the running of the  $\kappa_i$:
\begin{widetext}
\begin{eqnarray}
(\kappa_{2})_0 &=& \kappa_{2} +\frac13 \frac{q^2}{(4\pi)^2 \epsilon}\Big( 1 +13 \kappa_{2} + 4\kappa_{2} (\kappa_2 - 8\kappa_3 - 4\kappa_4) - 52 \kappa_{3} -36\kappa_{4} -8 \kappa_{5} +8 \kappa_{7}\Big) \label{Div1kappa2} \\[4pt]
(\kappa_{3})_0 &=& \kappa_{3} + \frac{1}{12}\frac{q^2}{(4\pi)^2 \epsilon}\Big(-1- \kappa_{2} +40 \kappa_{3}+ 16 \kappa_{3} (\kappa_2 - 8\kappa_3 - 4\kappa_4) - 4\kappa_{5} +16 \kappa_{7}\Big) \label{Div1kappa3} \\[4pt]
(\kappa_{4})_0 &=& \kappa_{4} +\frac16 \frac{q^2}{(4\pi)^2 \epsilon} \Big(-1 + 9 \kappa_{2} + 68 \kappa_{3} +38 \kappa_{4} +8 \kappa_{4} (\kappa_2 - 8\kappa_3 - 4\kappa_4) - 4\kappa_{5} -8 \kappa_{7}\Big) \label{Div1kappa4} \\[4pt]
(\kappa_{5})_0 &=& \kappa_{5} + \frac{1}{12}\frac{q^2}{(4\pi)^2 \epsilon} \Big(-3- 3\kappa_{2} -60 \kappa_{3} - 18 \kappa_{4}+ 12 \kappa_{5} +16 \kappa_{5} (\kappa_2 - 8\kappa_3 - 4\kappa_4) +24 \kappa_{7}\Big) \label{Div1kappa5}\\[4pt]
(\kappa_{6})_0 &=& \kappa_{6} +\frac{1}{36} \frac{q^2}{(4\pi)^2 \epsilon} \Big(-6-5 \kappa_{2} +132 \kappa_{3} -24\kappa_{4}- 24 \kappa_{5} +48 \kappa_{6} (\kappa_2 - 8\kappa_3 - 4\kappa_4) +24 \kappa_{7}\Big) \label{Div1kappa6}\\[4pt]
(\kappa_{7})_0 &=& \kappa_{7} +  \frac{1}{24}\frac{q^2}{(4\pi)^2 \epsilon}\Big(21+ 3\kappa_{2} -48 \kappa_{3} +6\kappa_{4} + 12 \kappa_{5} +32 \kappa_{7} (\kappa_2 - 8\kappa_3 - 4\kappa_4) \Big) \label{Div1kappa7}.
\end{eqnarray}
A somewhat unusual quadratic running follows from
renormalization of $\theta^{\mu\nu}$. In particular, we obtain
\begin{equation}
 (\kappa_{2})_0 -4 (\kappa_{3})_0 -2 (\kappa_{4})_0=(1+\frac43
 \frac{q^2}{(4\pi)^2 \epsilon})(\kappa_2 - 8\kappa_3 - 4\kappa_4) (\kappa_2 -4\kappa_3-2\kappa_4)+\frac 13
 \frac{q^2}{(4\pi)^2 \epsilon}(3+5\kappa_2 -160\kappa_3-74\kappa_4).  \nonumber
\end{equation}
\end{widetext}
We thus see that  constraint (\ref{relation}) is
not preserved for the bare couplings, that is, that coupling
constants $\kappa_2$, $\kappa_3$ and $\kappa_4$ do not renormalize
consistently with the SW expansion. Running of $\kappa_5$, $\kappa_6$ and $\kappa_7$
on the other hand obstructs  (\ref{**}-\ref{***}) not.

\section{Anomaly-safe theories}

The  conclusion is therefore that the SW-expanded
chiral electrodynamics (\ref{PolazniLag})  is not perturbatively renormalizable.
Of course, this theory is  not renormalizable for another,
stronger reason: the existence of the chiral anomaly.
It was namely  shown in \cite{Banerjee:2001un,Martin:2002nr} that
in the $\theta$-expanded  theories,  anomalies
and anomaly-cancellation conditions  are exactly
the same as in the corresponding commutative theories, and
know that  chiral electrodynamics contains the chiral anomaly.
So there is a natural the question: if we construct a model consisting of several
 fermions in different representations of noncommutative  $\mathrm{U}(1)$ 
and impose the anomaly-cancellation  conditions,  would
renormalizability  improve  as  it happens for the
GUT compatible models   discussed in \cite{Martin:2009sg,Martin:2009vg,Tamarit:2009iy}?
Unfortunately, as we shall see in the following, that this is not the case.
Though the anomaly-cancellation conditions
\begin{equation}
 \sum_i q_i =0, \qquad \qquad \sum_i q_i^3 =0
\end{equation}
remove the 3-photon vertex from the action and
make room  for an arbitrary renormalization of
 $\theta^{\mu\nu}$,  this additional  renormalization cannot improve
overall renormalizability of the model.

To show this let us shortly discuss the classical action.
We assume that we have a set of fermion fields
$\,  \hat\varphi_i$, $i=1,\dots N$ with electric charges  $q_i$.
It is known \cite{Aschieri:2002mc}
 that in the $\theta$-expanded theories each noncommutative
field comes with its own noncommutative potential $\hat A_i^\mu$;
all of them however for $\theta=0$ reduce  to the same
  $A^\mu$.  The $\hat A_i^\mu$ are different
because their corresponding SW maps  differ: they depend on charges
 $q_i$.  Therefore, in order to obtain the action
with the correct limit,  in the sum 
\begin{equation}
 \mathcal{L}_{\rm C}=\sum_{i=1}^N\bar\varphi_i \bar\sigma_\mu (\partial^\mu + {\rm i} q_i A^\mu_i)\varphi_i- \frac{1}{4} \sum_{i=1}^N F_i^{\mu\nu}F_{i,\mu\nu}
\end{equation}
we first need to rescale  gauge fields  $\, A^\mu_i \rightarrow \sqrt{c_i} A_i^\mu = \sqrt{c_i} A^\mu$
and charges $\, q_i \rightarrow \frac{1}{\sqrt{c_i}}q_i$. We then get
\begin{equation}
 \mathcal{L}_{\rm C}=\sum_{i=1}^N\bar\varphi_i \bar\sigma^\mu (\partial_\mu + {\rm i} q_i A_\mu)\varphi_i- \frac{1}{4} \sum_{i=1}^N c_i F^{\mu\nu}F_{\mu\nu}  .                     \label{lN}
\end{equation}
To associate (\ref{lN}) with the lagrangian of the usual commutative theory we 
fix the sum of weights $c_i$ to 1, $\ \sum_{i=1}^N c_i=1$. 
In the noncommutative $\mathrm{U}(1)$ case we are
free to choose  $c_i=1/N$; in other cases,  for example in 
noncommutative generalizations of the Standard Model, 
analogous relationsare more complicated, \cite{Buric:2006wm}.
The same rescaling applied to the noncommutative
 part of the lagrangian gives, for the boson vertex
\begin{equation}
 \mathcal{L}_{1,A} = \frac{\sum_i q_i}{2 N} \theta^{\mu\nu} \left( F_{\mu\rho}F_{\nu\sigma}F^{\rho\sigma} -\frac{1}{4} F_{\mu\nu}F_{\rho\sigma}F^{\rho\sigma}\right)    , \label{L1AN}
\end{equation}
and clearly in the anomaly-safe model in which
$\, \sum_i q_i=0\, $ we obtain that  this term vanishes,  $\mathcal{L}_{1,A}=0$.
Fermion terms are on the other hand
unchanged as they are mutually independent for each field:
\begin{widetext}
\begin{eqnarray}
\mathcal{L}_{2,\varphi_i} &=& \frac{\rm i}{16} \theta^{\mu\nu} \Delta^{\alpha\beta\gamma}_{\mu\nu\rho}\, F_{\alpha\beta} \,\sum_i q_i\bar\varphi_i\,
\bar\sigma^\rho (\partial_\gamma + {\rm i}q_i A_\gamma) \varphi_i \hc
\label{L1phiN}  \\[4pt]
\mathcal{L}_{3,\varphi_i}&=&{\mathrm{i}} \theta^{\mu\nu}\varepsilon_{\mu\nu\rho\sigma}\sum_i \bar\varphi_i \bar\sigma^{\sigma} D^{\rho} D^{2} \varphi_i\hc \nonumber  \\[4pt]
 \mathcal{L}_{4,\varphi_i}&=& \mathrm{i} \theta^{\mu\nu} F_{\mu\rho} \sum_i q_i \bar\varphi_i \bar \sigma_{\nu} D^{\rho}\varphi_i \hc   \nonumber\\[4pt]
\mathcal{L}_{5,\varphi_i}&=& \mathrm{i} \theta^{\mu\nu} F_{\mu\nu} \sum_i q_i \bar\varphi_i \bar \sigma^{\rho} D_{\rho}\varphi_i \hc  \nonumber  \\[4pt]
\mathcal{L}_{6,\varphi_i}&=&\theta^{\mu\nu} \varepsilon_{\mu\rho\sigma\tau}F^{\rho\sigma} \sum_i q_i\bar\varphi_i \bar \sigma^{\tau} D_{\nu}\varphi_i \hc  \nonumber \\[4pt]
\mathcal{L}_{7,\varphi_i}&=& \theta^{\mu\nu} \varepsilon_{\mu\nu\rho\sigma}F^{\rho\sigma} \sum_i q_i\bar\varphi_i \bar \sigma^{\tau} D_{\tau}\varphi_i \hc              .  \nonumber
\end{eqnarray}
\end{widetext}

We can extract the value of the one-loop divergences from our previous result either
using the same  rescaling of charges $q_i$ by $c_i$  or straightforwardly,
by repeating the  calculation. We obtain for the renormalized lagrangian:
\begin{widetext}
\begin{eqnarray}
{\mathcal{L}}^\prime_{\rm NC}+{\mathcal{L}}^\prime_{\rm ct}&=& - \frac{1}{4}F_{\mu\nu}F^{\mu\nu}\left(1-\frac{4}{3} \frac{\sum_i q_i^2}{(4\pi)^2 \epsilon}\right) +\sum_i\left( \frac{\mathrm{i}}{2}\bar\varphi_i\bar\sigma^\mu( D_\mu \varphi_i )\hc\right)  \left(1-2 \frac{q_i^2}{(4\pi)^2 \epsilon}\right) \\
& & +\frac{1}{16} \mu^{\frac{\epsilon}{2}} \theta^{\mu\nu} \Delta^{\alpha\beta\gamma}_{\mu\nu\rho}\, F_{\alpha\beta}\sum_i \left({\rm i}q_i\bar\varphi_i \bar\sigma^\rho ( D_\gamma \varphi_i) \hc\right)\left(1+ \kappa_{i,2} +\alpha_{i,2} \frac{q_i^2}{(4\pi)^2 \epsilon}   \right)  \nonumber\\
& & +\theta^{\mu\nu}\varepsilon_{\mu\nu\rho\sigma} \sum_i \left({\mathrm{i}} \bar\varphi_i \bar\sigma^{\sigma} D^{\rho} D^{2} \varphi_i\hc \right) \left( \kappa_{i,3} + \alpha_{i,3} \frac{q_i^2}{(4\pi)^2 \epsilon} \right)  \nonumber\\
& & +\mu^{\frac{\epsilon}{2}}\theta^{\mu\nu} F_{\mu\rho}\sum_i \left(\mathrm{i}q_i \bar\varphi_i \bar \sigma_{\nu} D^{\rho}\varphi_i \hc\right) \left( \kappa_{i,4} + \alpha_{i,4} \frac{q_i^2}{(4\pi)^2 \epsilon} \right) \nonumber \\
& & +\mu^{\frac{\epsilon}{2}}\theta^{\mu\nu} F_{\mu\nu} \sum_i \left( \mathrm{i}q_i\bar\varphi_i \bar \sigma^{\rho} D_{\rho}\varphi_i \hc \right) \left( \kappa_{i,5} + \alpha_{i,5} \frac{q_i^2}{(4\pi)^2 \epsilon} \right) \nonumber\\
& & +\mu^{\frac{\epsilon}{2}}\theta^{\mu\nu} \varepsilon_{\mu\rho\sigma\tau}F^{\rho\sigma} \sum_i \left(q_i\bar\varphi_i \bar \sigma^{\tau} D_{\nu}\varphi_i \hc \right) \left( \kappa_{i,6} + \alpha_{i,6} \frac{q_i^2}{(4\pi)^2 \epsilon}  \right) \nonumber\\
& & +\mu^{\frac{\epsilon}{2}}\theta^{\mu\nu} \varepsilon_{\mu\nu\rho\sigma}F^{\rho\sigma}\sum_i \left(q_i\bar\varphi_i \bar \sigma^{\tau} D_{\tau}\varphi_i \hc\right) \left( \kappa_{i,7} + \alpha_{i,7} \frac{q_i^2}{(4\pi)^2 \epsilon} \right), \nonumber
\end{eqnarray}
\end{widetext}
with
\begin{eqnarray*}
\alpha_{i,2}&=&\frac13(-5+3\kappa_{i,2} - 20 \kappa_{i,3} - 20 \kappa_{i,4} - 8\kappa_{i,5} + 8\kappa_{i,7}),\\
\alpha_{i,3}&=&\frac{1}{12}(-1 -\kappa_{i,2} + 16 \kappa_{i,3} +14 \kappa_{i,4} -4\kappa_{i,5} +16\kappa_{i,7}),\\
\alpha_{i,4}&=&\frac16(-1 + 9\kappa_{i,2}+ 68 \kappa_{i,3} + 26 \kappa_{i,4} - 4\kappa_{i,5} - 8 \kappa_{i,7}),\\
\alpha_{i,5}&=&\frac14(-1 - \kappa_{i,2} - 20 \kappa_{i,3} -6\kappa_{i,4} - 4 \kappa_{i,5} + 8 \kappa_{i,7}),\\
\alpha_{i,6}&=&\frac{1}{36}(-6 -5 \kappa_{i,2}+ 132 \kappa_{i,3} -24 \kappa_{i,4}- 24 \kappa_{i,5} \\
&& - 72 \kappa_{i,6} + 24 \kappa_{i,7}),\\
\alpha_{i,7}&=&\frac18(7 +\kappa_{i,2} -16\kappa_{i,3} + 2\kappa_{i,4} + 4\kappa_{i,5} - 16 \kappa_{i,7}).
\end{eqnarray*}
Renormalization of  fields and charges is the standard one,
\begin{eqnarray*}
\varphi_{i,0}&=&\sqrt{Z_{i,2}}\varphi_i=\sqrt{1-2 \frac{q_i^2}{(4\pi)^2 \epsilon}} \varphi_i,
\\
A^\mu_0&=&\sqrt{Z_3} A^\mu=\sqrt{1-\frac{4}{3} \frac{\sum_j q_j^2}{(4\pi)^2 \epsilon}} A^\mu,\\
q_{i,0}&=&\mu^{\frac{\epsilon}{2}} Z_3^{-1/2}Z_{i,2}^{-1}   \left(1-2 \frac{q_i^2}{(4\pi)^2 \epsilon}\right) q_i \\
&=& \mu^{\frac{\epsilon}{2}} \left(1+\frac{2}{3} \frac{\sum_j q_j^2}{(4\pi)^2 \epsilon}\right) q_i ,
\end{eqnarray*}
while noncommutativity $\theta^{\mu\nu}$ can renormalize arbitrarily. In order to
try to use this fact we assume that it is  of the form
$$
\theta^{\mu\nu}_0=\left(1+\alpha\, \frac{\sum_j q_j^2}{(4\pi)^2 \epsilon} \right) \theta^{\mu\nu},
$$
with an arbitrary coefficient $\alpha $ which is  to be determined from some renormalizability constraint.
Renormalization of the $\,\kappa_i$ follows:
\begin{widetext}
\begin{eqnarray*}
(\kappa_{i,2})_0 &=& \kappa_{i,2} + \frac{1}{(4\pi)^2 \epsilon}\left( -\alpha (1+\kappa_{i,2}) \sum_j q_j^2 + \frac13  (1 + 9\kappa_{i,2} - 20\kappa_{i,3} -20\kappa_{i,4} -8 \kappa_{i,5} +8 \kappa_{i,7})q^2_i  \right), \label{Divkappa2} \\
(\kappa_{i,3})_0 &=& \kappa_{i,3} + \frac{1}{(4\pi)^2 \epsilon} \left(-\alpha \kappa_{i,3} \sum_j q_j^2 + \frac{1}{12}(-1- \kappa_{i,2} +40 \kappa_{i,3} +14 \kappa_{i,4} - 4\kappa_{i,5} +16 \kappa_{i,7})q^2_i  \right),   \label{Divkappa3} \\
(\kappa_{i,4})_0 &=& \kappa_{i,4} + \frac{1}{(4\pi)^2 \epsilon} \left(-\alpha\kappa_{i,4} \sum_j q_j^2 + \frac16 (-1+9 \kappa_{i,2} +68 \kappa_{i,3} +38 \kappa_{i,4} - 4\kappa_{i,5} -8 \kappa_{i,7})q^2_i   \right),  \label{Divkappa4} \\
(\kappa_{i,5})_0 &=& \kappa_{i,5} + \frac{1}{(4\pi)^2 \epsilon} \left(-\alpha \kappa_{i,5} \sum_j q_j^2 + \frac14(-1- \kappa_{i,2} -20 \kappa_{i,3} -6\kappa_{i,4}+ 4 \kappa_{i,5} +8 \kappa_{i,7})q^2_i \right),   \label{Divkappa5}\\
(\kappa_{i,6})_0 &=& \kappa_{i,6} + \frac{1}{(4\pi)^2 \epsilon} \left(-\alpha \kappa_{i,6} \sum_j q_j^2 + \frac{1}{36}(-6-5 \kappa_{i,2} +132 \kappa_{i,3} -24\kappa_{i,4}- 24 \kappa_{i,5} +24 \kappa_{i,7})q^2_i \right),   \label{Divkappa6}\\
(\kappa_{i,7})_0 &=& \kappa_{i,7} + \frac{1}{(4\pi)^2 \epsilon} \left(-\alpha \kappa_{i,7} \sum_j q_j^2 + \frac18(7+ \kappa_{i,2} -16 \kappa_{i,3} +2\kappa_{i,4} + 4 \kappa_{i,5} ) q^2_i \right)  \label{Divkappa7}   .
\end{eqnarray*}
But we easily observe that, however we fix $\alpha$,  expressions
\begin{eqnarray*}
(\kappa_{i,2})_0 -4 (\kappa_{i,3})_0 -2(\kappa_{i,4})_0& =&
\left(  1-  \frac{1}{(4\pi)^2 \epsilon} \alpha\sum_j q_j^2 \right) (\kappa_{i,2}-4 \kappa_{i,3} -2\kappa_{i,4})
-   \frac{1}{(4\pi)^2 \epsilon} \alpha\sum_j q_j^2 \\[4pt]
&& + \frac{1}{(4\pi)^2 \epsilon}\frac{q_i}{3}( 3+19\kappa_{i,2} -128 \kappa_{i,3}-72\kappa_{i,4}),
\end{eqnarray*}
cannot be zero.
\end{widetext}

\section{Discussion}

Before we discuss the meaning  of our result let us consider briefly
some limiting cases. The easiest case is when fermions are absent,
$\, \varphi=0$. We see  that  then there  are no new noncommutative divergences
 and therefore no need to rescale $\theta^{\mu\nu}$, which  is
in accord with previously obtained
behavior of the $SU(N)$ gauge theories, \cite{Buric:2005xe}. We discussed in \cite{Buric:2010wd}
 the minimal or `little' case  in which all  $\kappa_i=0$, $i=2,\dots,7$, that is, in which 
complete noncommutative fermion correction reduces to $ \mathcal{L}_{2}$. 
As it can be seen from
(\ref{Div1kappa2}-\ref{Div1kappa7}),  this one classical term
generates after quantization all six $\mathcal{L}_{i} $; even the
fermion propagation changes. This is in a way  nice result, as it
shows again a specific relation between the spatial and the gauge degrees
 of freedom: fermion propagation changes because of the photon loops.
Quantum corrections  generate noncommutative interactions even when
they are absent from the classical lagrangian, that is for
 $1+\kappa_2=0$, $\kappa_i=0$, $i=3,\dots,7$;
this effect was discussed before in the case of Dirac fermions in
\cite{Grimstrup:2002af}.

Let us summarize the  obtained result.
We started classically with the most general
action permitted by the Seiberg-Witten map. As the amount of
nonuniqueness of the Seiberg-Witten expansion is huge, the initial
lagrangian  (\ref{PolazniLag}) contains essentially
all terms allowed by dimension and gauge covariance.
Denote
\begin{equation}
  \mathcal{L}_{1,A}^\prime =
\lambda_1\theta^{\mu\nu} F_{\mu\rho}F_{\nu\sigma}F^{\rho\sigma} +
\lambda_2\theta^{\mu\nu}F_{\mu\nu}F_{\rho\sigma}F^{\rho\sigma} .
\end{equation}
There are only two conditions in the lagrangian which distinguish the
origin of the separate terms, that is which signify
that the action was derived from (\ref{act}) through the
SW map: they are
\begin{equation}
 \lambda_1+4\lambda_2 =0,\qquad
 \kappa_2-4\kappa_3-2\kappa_4 =0              .
\end{equation}
The first relation, the ratio between the two 3-photon
terms, is stable under quantization; the second relation is broken after
renormalization of the theory. This means that the SW map
is not compatible  with quantization: clearly, this happens
only when fermions are present. This implies that the $\theta$-expanded
chiral electrodynamics is not renormalizable, and we are
forced to conclude more generally,  that the $\theta$-expanded theories
cannot be considered as fundamental or basic theories which provide
 representations of gauge symmetry on the Moyal space. They
can probably give a good effective description of the effects of
noncommutativity, but we expect that  in a fundamental noncommutative
gauge theory matter will have to be included in a different way.

\begin{acknowledgments}
This work was supported by the Serbian Ministry of Education, Science and
Technological Development under Grant No. ON 171031.
\end{acknowledgments}

\end{document}